# AN ALTERNATE RING-RING DESIGN FOR eRHIC

Yuhong Zhang*

Thomas Jefferson National Accelerator Facility, Newport News, VA 23606, USA

*Abstract*

I present here a new ring-ring design of eRHIC, a polarized electron-ion collider based on RHIC at BNL. This alternate eRHIC design utilizes high repetition rate colliding beams and is likely able to deliver the performance to meet the requirements of the science program with low technical risk and modest accelerator R&D. The expected performance includes high luminosities over multiple collision points and a broad CM energy range with a maximum value up to $2\times10^{34}$ cm$^{-2}$s$^{-1}$ per detector, and polarization higher than 70% for the colliding electron and light ion beams. This new design calls for reuse of decommissioned facilities in the US, namely, the PEP-II high energy ring and one section of the SLAC warm linac as a full energy electron injector.

## INTRODUCTION

A polarized electron-ion collider (EIC) in a CM energy range up to 100 GeV/n is envisioned as a future facility for nuclear science research in the US [1]. Presently, both BNL and Jefferson Lab are engaged in design studies for this future collider [2,3,4,5]. Since the beginning in 2001, the EIC design studies have been focused on achieving superior performance. The EIC science program [1] requires high luminosity (above $10^{33}$ cm$^{-2}$s$^{-1}$ per detector over a broad CM energy range with a wide array of fully stripped ion species up to lead or uranium), and high polarization (>70%) for both colliding electron and light ion beams. Ring-ring and ERL-ring collider scenarios have been adopted respectively for MEIC (the Jefferson Lab design [4,5]) and eRHIC (the BNL design [3]).

The MEIC design is based on a very high bunch repetition rate for both colliding beams and strong final focusing to achieve a luminosity close to $10^{34}$ cm$^{-2}$s$^{-1}$ per detector [6]. The design concept was formulated more than 10 years ago and is still considered the best approach for MEIC. The design adopted a figure-8 shape for the ion booster and both collider rings to achieve high beam polarization for all polarized ion species including deuterons [4,5]. Recently, the reuse of the PEP-II high energy ring for the MEIC electron collider ring has been integrated into the present baseline [5].

eRHIC started with a ring-ring collider design [2] and advanced to the present ERL-ring collider design [3] around 2007 to maximize luminosity. This design is very innovative and introduces a set of new and advanced concepts and schemes, and relies on several yet-to-be-demonstrated accelerator technologies [3]. These include a high current polarized electron source based on a multi-cathode Gatling gun; high energy high current multi-pass ERL based on two sets of FFAG-like recirculation beam lines; space charge compensation; and coherent electron cooling. While these concepts, schemes and technologies are expected to improve the eRHIC performance, they do require substantial R&D effort for development and proto-typing, and for proof-of-principle demonstrations.

In this report I propose an alternate eRHIC design based on a ring-ring collider scenario and the same luminosity concept used in the MEIC design. This design has the potential to reach high luminosities and polarization while requiring modest accelerator R&D. I also include a discussion on several implementation issues and the required accelerator R&D.

## DESIGN CONCEPT

### Design Strategy

I choose a ring-ring collider scenario as the basis of this alternate eRHIC design since it is technically a conservative approach compared to an ERL-ring design but still able to deliver very high luminosities.

The ring-ring eRHIC design supports simultaneous operation of multiple detectors, and thus increases science productivity. This is different from an ERL-ring design where an electron bunch is allowed to collide only once (at one of the detectors) due to a large beam-beam disruption, so the detectors can only be operated alternately.

The strategy for high luminosity has been demonstrated already in the B-factories [6] and has been adopted for the MEIC design since 2002 [4]. The concept involves specific choices in the design of colliding beams, interaction region and beam cooling. Namely, (1) both colliding beams have a high bunch repetition rate, very small bunch lengths, bunch charges and transverse emittances; (2) the interaction region has very strong final focusing to attain very small beam spots at collision points; it also has an implementation of crab crossing of colliding beams to support high bunch repetition; (3) electron cooling is responsible for reduction of the proton or ion beam six-dimensional emittances.

I further propose to reuse the PEP-II high energy ring and part of the SLAC warm linac for eRHIC. Following this approach, no major new facility is required.

### New Design Baseline

The following key elements are proposed as the new ring-ring eRHIC baseline

- *Electron collider ring*: the PEP-II high energy ring;
- *Ion collider ring*: one of the RHIC rings;
- *Electron full energy injector*: a section of the SLAC warm linac;
- *Ion injector*: the RHIC ion complex

Reuse of the PEP-II high energy ring includes the entire magnet set and the vacuum chamber as well as the RF system. This results in a 476 MHz bunch repetition rate of the electron beam in the collider ring.

### RHIC Upgrade

___________________________________________

\* yzhang@jlab.org

I further propose to upgrade RHIC and its injector complex
- To support high bunch repetition rates;
- To support multi-phased electron cooling;
- To improve the RHIC polarized ion beam operation.

To match the high bunch repetition rate of the electron beam, the RF system of RHIC (and part of AGS) should be rebuilt to support 476 MHz frequency. This provides an opportunity to implement the high luminosity concept based on a high bunch repetition rate discussed above.

Besides a new RF system, the RHIC upgrade needs two electron cooling facilities, namely, a DC cooler installed in the AGS and a bunched beam ERL based cooler installed in the RHIC ring. More Siberian snakes will be installed in the RHIC ring as already planned in the present eRHIC design [3].

## IMPLEMENTATION ISSUES

### PEP-II Ring in the RHIC Tunnel

Both RHIC and PEP-II rings have six-fold symmetric footprints, however, their circumferences are different, namely, 3834 m for RHIC and 2199 m for PEP-II [7]. There are two ways to fit the PEP-II high energy ring into the RHIC tunnel. The first way is by reducing the dipole packing factor of the arc cells to stretch the PEP-II arc length while preserving the dipole bending angle, therefore, no new magnet is required. The second way is by adding extra PEP-II type arc cells to cover the RHIC arc. This will reduce the dipole bending angle and increase the dipole bend radius, leading to reduction of synchrotron radiation (SR) power. Table 1 lists the number of electron arc cells in each sextant (~399 m) of the RHIC ring, the corresponding dipole bending radius and SR power using these two methods. The PEP-II vacuum chambers have a limit of 10 kW/m for the SR power density. Thus in the high energy region, the electron current must be adjusted to satisfy this limit.

**Table 1**: Fitting the PEP-II arc cells into one sextant of the RHIC tunnel; the third row shows stretched PEP-II arc cells; the fourth row utilizes more PEP-II type arc cells.

| Cells per arc | Packing factor | Bending radius | Cell length | Current @10GeV | SR power |
|---|---|---|---|---|---|
| | | m | m | A | MW |
| 16 | 0.43 | 165 | 24.9 | 1.93 | 10 |
| 26 | 0.71 | 268 | 15.2 | 3 | 5.9 |

### Injection from the SLAC Linac

The SLAC warm linac was used as a full energy injector to the PEP-II collider rings. The currents of stored high energy electron and low energy positron beams reached 2.1 A and 3.2 A respectively during the last run of the PEP-II program [8]. To support eRHIC, a pulsed polarized electron source is required. To match the PEP-II injection performance, this source should be able to deliver a beam with a 60 to 120 Hz pulse repetition rate, (0.1 to 3)×$10^{10}$ electrons per pulse, and polarization larger than 70%. Such a source was developed at SLAC for Stanford Linear Collider (SLC) operations in the early 90s [9]. It delivered 7×$10^{10}$ electrons per pulse, and had a 4-day QE lifetime. With this source and the SLAC pulsed linac, the eRHIC electron ring can be filled in a similar time as the PEP-II rings, as low as 2 minutes.

Top-off injection was performed in PEP-II in 30 min interval and took 3 minutes to bring the stored current from 80% to 100%. Such a top-off injection scheme is useful and also sufficient to maintain a satisfactory beam current and polarization in eRHIC. Since the synchrotron radiation is now low due to a larger dipole bending radius, the depolarization time is on the order of hours even at very high energies.

If it is decided that the SLAC linac will not be used for eRHIC, a new warm linac may be considered. Recirculation of up to three to four passes of the linac may reduce the linac energy substantially.

## COLLIDER PERFORMANCE

The main features of this alternate ring-ring eRHIC design can be summarized below
- Multiple detectors and simultaneous operation;
- Ion species: proton to uranium, fully stripped; proton and helium-3 ions are polarized;
- Electron energy range: 3 to 15 GeV;
- Proton energy range: up to 250 GeV;
- Ion energy range: up to 100 GeV/u
- Center-of-mass energy: 15.8 to 122.5 GeV/u;
- Electron beam current: 3 A nominal value, reduced at high energies;
- Proton beam current: 0.415 A nominal value, same as the RHIC operation current
- Bunch repetition rate of colliding beams: 476 MHz nominal value;
- Protons per bunch: 5.4×$10^9$ nominal value;
- Electron polarization: >70%, with longitudinal polarization at collision points
- Proton and helium-3 polarization: ~70%, with both longitudinal or transverse polarization at collision points
- Peak luminosity is up to 2×$10^{34}$ $cm^{-2}s^{-1}$ per detector for a full acceptance detector

The PEP-II high energy ring, which is inherited from the PEP project, can support 18 GeV electrons [7]. If more PEP-II like arc cells are packed into the RHIC tunnel, the ring can accommodate an electron beam with energy up to ~30 GeV.

The SLAC warm linac has a total energy of 50 GeV. It is divided into three sections, with one section presently used as a driving linac for the LCLS X-FEL source; the second section is used for experimental research of plasma acceleration of charged particles. The third section of this linac is scheduled to be dismantled to make the space available for a new SRF linac for the LCLS-II project. This section can accelerate electron at least to 15 GeV. Studies showed that this section of the linac can reach 30 GeV by installing twice many klystrons.

The nominal electron current for this ring-ring eRHIC design is 3 A; however, it must be scaled down proportionally for beam energies above 9 GeV if only the PEP-II magnets are used, in order to meet the SR power density limit (10 kW/m) of the PEP-II vacuum chambers [7]. On the other hand, if more PEP-II like arc cells are used to fill the RHIC tunnel, the SR power is reduced due to a larger dipole bending radius, and a 3 A nominal current can be maintained up to 11.2 GeV beam energy.

The nominal current of the eRHIC proton beam is set to 0.415 A, which has already been achieved in RHIC operations. Due to the high bunch repetition rate (up to 476 MHz), the proton bunch charge is very small, i.e., $5.4 \times 10^9$ protons per bunch, which is roughly 56 times smaller than that of the ERL-ring eRHIC design ($3 \times 10^{11}$ protons per bunch [3]).

Table 2 summarizes main parameters of $e$-$p$ collisions for this alternate ring-ring eRHIC design at four representative energy points. The electron current and emittance are estimated for the case that the collider ring is filled with extra PEP-II type arc cells. The luminosities are above $10^{33}$ cm$^{-2}$s$^{-1}$ per detector at all four design points, and the corresponding CM energy is from 31.6 to 122.5 GeV. The maximum luminosity is $2 \times 10^{34}$ cm$^{-2}$s$^{-1}$ at a medium CM energy (71 GeV). The luminosities of $e$-$A$ collisions can be estimated similarly, and they reach a similar high level. Figure 1 shows the luminosity curve for the $e$-$p$ collisions.

**Table 2**: Main design parameters of the new EIC design at four representative design points.

| CM energy | GeV | 31.6 (low) | | 70.7 (medium) | | 100 (high) | | 122.5 (high) | |
|---|---|---|---|---|---|---|---|---|---|
| | | $p$ | $e$ | $p$ | $e$ | $p$ | $e$ | $p$ | $e$ |
| Beam energy | GeV | 50 | 5 | 250 | 5 | 250 | 10 | 250 | 15 |
| Bunch repetition rate | MHz | 476 | | 476 | | 476 | | 476/3=158.7 | |
| Particles per bunch | $10^{10}$ | 0.54 | 3.9 | 0.54 | 3.9 | 0.54 | 3.9 | 1.6 | 3.9 |
| Beam current | A | 0.415 | 3 | 0.415 | 3 | 0.415 | 3 | 0.415 | 1 |
| Polarization | % | 70 | 70 | 70 | 70 | 70 | 70 | 70 | 70 |
| Bunch length, RMS | cm | 2 | 1.2 | 1 | 1.2 | 1 | 1.2 | 2 | 1.2 |
| Norm. emit., vert./horz. | μm | 0.5 | 11.2 | 0.5 | 11.2 | 0.5 | 90 | 0.5 | 310 |
| Horizontal & vertical β* | cm | 2 | 16.5 | 2 | 3.3 | 2 | 1.5 | 3 | 1.35 |
| Vert. beam-beam parameter | | 0.01 | 0.11 | 0.01 | 0.11 | 0.005 | 0.025 | 0.004 | 0.03 |
| Laslett tune-shift | | 0.036 | Small | 0.003 | Small | 0.003 | Small | 0.004 | Small |
| Hour-glass (HG) factor | | 0.88 | | 0.92 | | 0.85 | | 0.74 | |
| Luminosity/IP, w/HG, $10^{33}$ | cm$^{-2}$s$^{-1}$ | 3.8 | | 19.8 | | 13.0 | | 6.0 | |

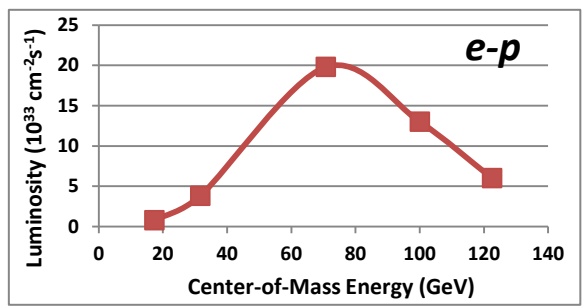

**Figure 1:** Luminosity of $e$-$p$ collisions of the alternative ring-ring eRHIC design.

In the parameter table above, the proton bunch length is 1 to 2 cm and the betatron function at the collision point (β*) is 2 to 3 cm, thus the luminosity reduction due to the hour glass effect is modest. An interaction region design supporting this ultra small β* and a full acceptance detector has been developed by the Jefferson Lab MEIC team [5]. Because of the very small bunch charges, the proton beam space charge tune-shift is very small (less than 0.01) in the medium to high energy range. Even at 50 GeV, the proton beam space charge tune-shift can be kept modest (0.036) after the bunch length is increased from 1 cm to 2 cm. At high electron energy (15 GeV), the electron current is reduced to 1 A. In this case, the luminosity is optimized by reducing the bunch repetition rate (and the collision frequency) by a factor of 3 to boost the bunch charge.

## ACCELERATOR R&D

Choosing a ring-ring design for eRHIC avoids several accelerator R&D efforts associated with the present ERL-ring eRHIC design. They include a 50 mA polarized electron source; high energy, high current, multi-path ERL and linac-ring beam-beam effect.

Due to a very small proton or ion bunch charge and modest emittances, the space charge effect is modest so space charge compensation [3] is not required. Similarly, the resistive wall effect is also much weaker than that of the present eRHIC design, and there is likely no need to upgrade the RHIC vacuum chambers (namely, coating the chambers) [3].

It is well understood that an efficient cooling mechanism must be a part of an EIC design for reaching its high luminosity goal. Since the proposed emittance (~0.5 mm mrad) of the colliding proton or ion beams is modest, the intra-beam scattering (IBS) time is not extremely short. As a consequence, conventional electron cooling with a multi-step cooling scheme [10,11] is likely sufficient for delivering and preserving the designed beam emittances and for achieving high luminosities. This cooling scheme utilizes a DC electron cooler in the AGS for an initial cooling and a bunched beam cooler in the

RHIC for final cooling and for suppressing the IBS induced emittance growth during collisions. The increased IBS time relaxes the current requirement in the bunched beam cooler to a few hundred mA, close to the present state-of-art, and therefore, success of the required electron source R&D is very likely. Strong beam cooling methods, such as coherent electron cooling [3] and bunched beam cooling driven by an ERL and a circulator ring [5], are currently under development at both BNL and Jefferson Lab. These methods could further reduce the proton and ion beam emittances to reach even higher luminosities.

## CONCLUSION

It seems that a ring-ring eRHIC can be constructed by reusing the existing RHIC and decommissioned facilities (PEP-II and SLAC linac). The alternative eRHIC design holds the promise to deliver excellent performance and to require modest accelerator R&D. An extra advantage is that the performance goals (namely, the CM energy range and luminosity) set by the EIC science program [1] can be met without requiring a staged approach.

## ACKNOWLEDGMENT

I am in debt to both the BNL eRHIC team and Jefferson Lab MEIC team for their outstanding design studies and accelerator R&D activities over the last 14 years. These works have greatly improved my understanding of the EIC proposals and form the foundation of the study presented in this paper.

I particularly thank Ya. Derbenev and R. Rimmer of Jefferson Lab, and M. Sullivan of SLAC for helpful discussions.

I also thank A. Hutton and R. Rimmer of Jefferson Lab for first suggesting reuse of the PEP-II magnets and vacuum chambers, and of RF systems respectively for MEIC.

Finally I want to thank the Jefferson Lab leadership for their support.

The work is authored by Jefferson Science Associates, LLC under U.S. DOE Contract No. DE-AC05-06OR23177 and DE-AC02-06CH11357. The U.S. Government retains a non-exclusive, paid-up, irrevocable, world-wide license to publish or reproduce this manuscript for U.S. Government purposes.